\documentclass[pdflatex,sn-nature]{sn-jnl}

\usepackage{graphicx}%
\usepackage{multirow}%
\usepackage{amsmath,amssymb,amsfonts}%
\usepackage{amsthm}%
\usepackage{mathrsfs}%
\usepackage[title]{appendix}%
\usepackage{xcolor}%
\usepackage{textcomp}%
\usepackage{manyfoot}%
\usepackage{booktabs}%
\usepackage{algorithm}%
\usepackage{algorithmicx}%
\usepackage{algpseudocode}%
\usepackage{listings}%
\usepackage{caption}

\theoremstyle{thmstyleone}%

\theoremstyle{thmstyletwo}%

\theoremstyle{thmstylethree}%

\AtBeginDocument{\fontsize{11}{13}\selectfont} 

\makeatletter
\renewcommand{\abstractfont}{\reset@font\fontsize{11bp}{13bp}\selectfont\leftskip=24pt\rightskip=24pt\parfillskip=0pt plus 1fil}
\renewcommand{\figurecaptionfont}{\reset@font\fontfamily{\rmdefault}\fontsize{10}{12}\selectfont}
\renewcommand{\abstractheadfont}{%
  \reset@font\fontsize{11}{13}\bfseries\selectfont\titraggedcenter
}

\makeatother

\begin{document}

\title[Article Title]{Nonvolatile Switching of Magnetism via Gate-Induced Sliding in Tetralayer Graphene} 

\author[1]{\fnm{Daniel} \sur{Brandon}}
 
\author[2]{\fnm{Tixuan} \sur{Tan}}

\author[3]{\fnm{Yiwen} \sur{Ai}}

\author[3]{\fnm{Peter} \sur{Golemis}} 

\author[3]{\fnm{Akshat} \sur{Gandhi}}

\author[3]{\fnm{Lujin} \sur{Min}}

\author[4]{\fnm{Kenji} \sur{Watanabe}}

\author[5]{\fnm{Takashi} \sur{Taniguchi}}

\author[2]{\fnm{Trithep} \sur{Devakul}}

\author*[3]{\fnm{Kenji} \sur{Yasuda}}\email{kenji.yasuda@cornell.edu}

\affil[1]{\orgdiv{Department of Physics}, \orgname{Cornell University}, \city{Ithaca}, \postcode{14853}, \state{NY}, \country{USA}}

\affil[2]{\orgdiv{Department of Physics}, \orgname{Stanford University}, \city{Stanford}, \postcode{94305}, \state{CA}, \country{USA}}

\affil[3]{\orgdiv{Department of Applied and Engineering Physics}, \orgname{Cornell University}, \city{Ithaca}, \postcode{14853}, \state{NY}, \country{USA}}

\affil[4]{\orgdiv{Research Center for Electronic and Optical Materials}, \orgname{National Institute for Materials Science}, \city{Ibaraki}, \postcode{305-0044}, \state{Tsukuba}, \country{Japan}}

\affil[5]{\orgdiv{Research Center for Materials Nanoarchitectonics}, \orgname{National Institute for Materials Science}, \city{Ibaraki}, \postcode{305-0044}, \state{Tsukuba}, \country{Japan}}
\abstract{Interlayer sliding degrees of freedom often determine the physical properties of two-dimensional (2D) materials \cite{li_binary_2017, fei_ferroelectric_2018, li_pressure-controlled_2019,  song_switching_2019, klein_enhancement_2019, yasuda_stacking-engineered_2021, vizner_stern_interfacial_2021, wang_interfacial_2022, weston_interfacial_2022, ko_operando_2023,   jindal_coupled_2023, vizner_stern_sliding_2025, liang_nanosecond_2025}. In graphene, for instance, the metastable rhombohedral stacking arrangement hosts correlated and topological electronic phases \cite{lee_competition_2014, shi_electronic_2020, chen_tunable_2020, zhou_half-_2021, zhou_superconductivity_2021, han_orbital_2023, liu_spontaneous_2024, lu_fractional_2024, choi_superconductivity_2025,  han_signatures_2025, morissette_superconductivity_2025, nguyen_hierarchy_2025, xie_tunable_2025}, which are absent in conventional Bernal stacking. Here, we demonstrate a sliding-induced first-order structural phase transition between Bernal and rhombohedral tetralayer graphene driven by gate voltages \cite{yankowitz_electric_2014, li_global_2020}. Through transport measurement, we observe bistable switching between a Bernal-dominant state and a rhombohedral-Bernal mixed state across a wide space of the gate-voltage phase diagram. The structural phase transition results in nonvolatile switching between a paramagnet and a ferromagnet accompanied by the anomalous Hall effect. The sign reversal of the anomalous Hall effect under opposite displacement fields suggests that it may originate from domain boundaries between the Bernal and rhombohedral regions. Our discovery paves the way for on-demand toggling of quantum phases based on the sliding phase transition of 2D materials and offers a playground to explore unconventional physics at the stacking domain boundaries.}

\maketitle

\section*{Main}\label{sec1}

First-order structural phase transitions enable nonvolatile control of the physical properties. Conventional approaches to induce a structural phase transition of a crystal rely on thermal, mechanical, chemical, or electrochemical stimuli, which are typically slow and energy intensive \cite{raoux_phase_2009, nakano_collective_2012, jeong_suppression_2013, jiang_manipulation_2018}. Electrostatic, gate-voltage-induced control of crystal structures in solid-state devices offers the potential for rapid, efficient, reversible, and reconfigurable modulation of material properties. 

This vision is hindered by the robustness of chemical bonds, whose energy scales are many orders of magnitude larger than what can be accessed through gate voltages. To address this challenge, we focus on two-dimensional (2D) materials, in which atomic planes are held together by weak van der Waals interactions, making them uniquely susceptible to external stimuli. The emerging paradigm of slidetronics demonstrates that interlayer sliding in 2D materials can indeed be driven by a gate voltage \cite{vizner_stern_sliding_2025}. For instance, in parallel-stacked bilayer hexagonal boron nitride (hBN), interlayer charge transfer induces a ferroelectric state, allowing a gate voltage to switch its polarization through interlayer sliding \cite{li_binary_2017, yasuda_stacking-engineered_2021, vizner_stern_interfacial_2021}. 

\begin{figure*}[h]
\centering
\includegraphics[width=1\textwidth]{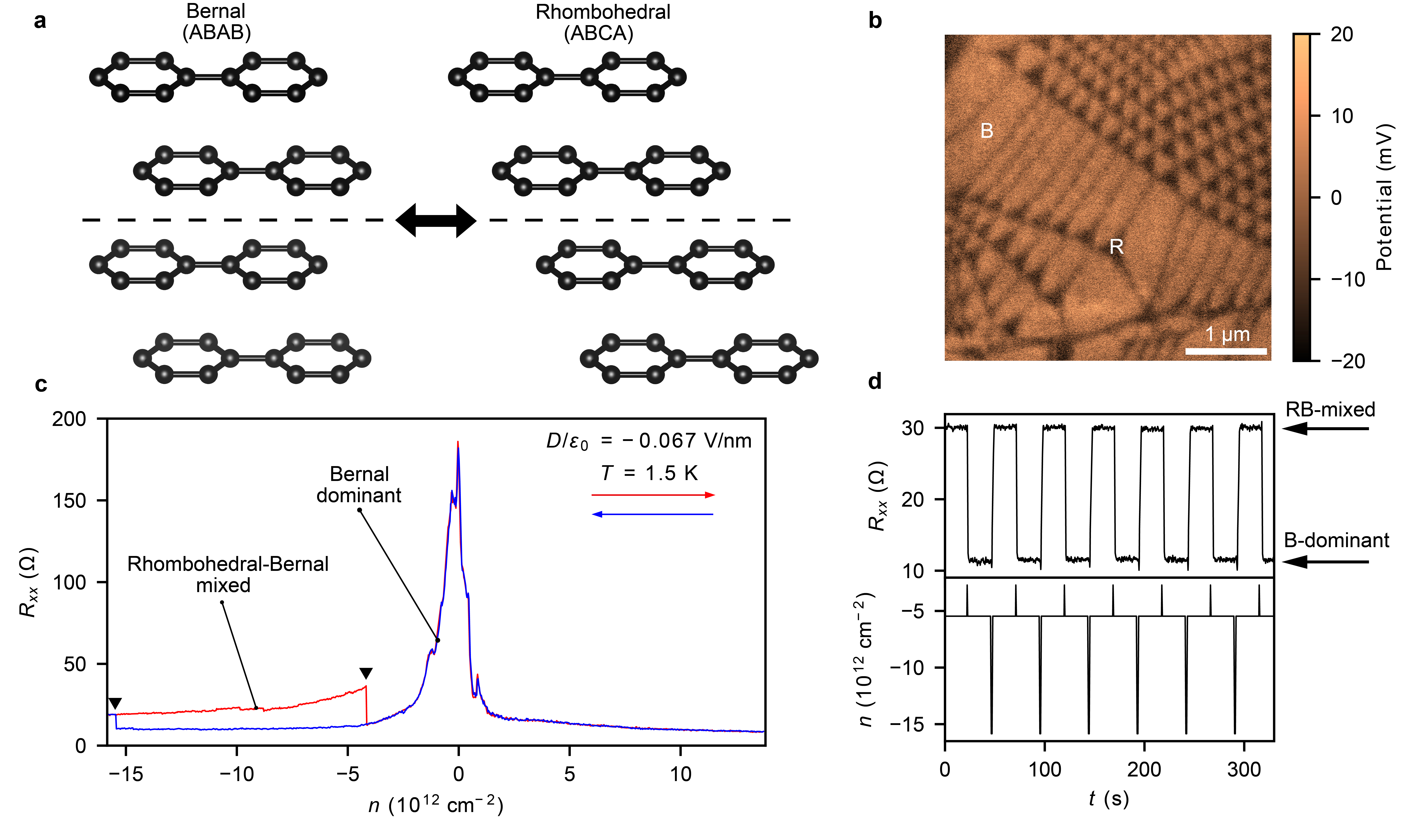}
\caption{\textbf{Doping-induced sliding phase transition between rhombohedral and Bernal tetralayer graphene.} \textbf{a,} Crystal structures of Bernal (ABAB) and rhombohedral (ABCA) tetralayer graphene drawn with VESTA \cite{momma_vesta_2011}. The two structures are related by an interlayer sliding between the second and the third layers of graphene. \textbf{b,} Kelvin probe force microscopy (KPFM) image of mixed Bernal (B, light) and rhombohedral (R, dark) domains. The Bernal regions is larger in area than the rhombohedral regions because of the lower energy of Bernal stacking. \textbf{c,} Longitudinal resistance $R_{xx}$ measured versus forward (red) and backward (blue) scan directions of carrier density $n$ at a constant displacement field of $D/\varepsilon_0 = -0.067  \ \mathrm{V/nm}$, where $\varepsilon_0$ is the vacuum permittivity. Abrupt changes in the resistance appear at $-4.1 \times 10^{12}\ \mathrm{cm}^{-2}$ and $-15.5 \times 10^{12}\ \mathrm{cm}^{-2}$ for the forward and backward scans, respectively, as highlighted by the black triangles. \textbf{d,} (Top) $R_{xx}$ measured versus time $t$. The high and low resistance regions correspond to the RB-mixed and B-dominant states, respectively. (Bottom) $n$ versus $t$ controlled under $D/\varepsilon_0 = -0.067  \ \mathrm{V/nm}$.}
\label{fig1}
\end{figure*}

Here, we extend the concept of slidetronics to realize a first-order quantum phase transition through gate-induced interlayer sliding motion. As a prototypical example, we investigate the sliding polymorphs of tetralayer graphene, i.e., Bernal (ABAB) and rhombohedral (ABCA) stacking orders (Fig. \ref{fig1}a). Rhombohedral graphite is of particular interest due to its topological flat bands \cite{guinea_electronic_2006, min_chiral_2008, koshino_trigonal_2009}, which host valley–spin-polarized magnetism \cite{zhou_half-_2021, han_orbital_2023, liu_spontaneous_2024}, integer and fractional quantum anomalous Hall states \cite{chen_tunable_2020, lu_fractional_2024, xie_tunable_2025}, and unconventional superconductivity \cite{zhou_superconductivity_2021,choi_superconductivity_2025,han_signatures_2025,morissette_superconductivity_2025,nguyen_hierarchy_2025}. Controlling the structural transition between Bernal and rhombohedral stacking, therefore, may enable switching of these intriguing quantum phases. 

In tetralayer graphene, sliding motion between the second and third layers can drive such a phase transition, as illustrated in Fig. \ref{fig1}a. Real-space imaging studies have shown that the sliding transition between the two stacking orders can be induced by electrostatic doping \cite{yankowitz_electric_2014, li_global_2020}. In this work, we will achieve this gate-induced sliding phase transition in low-temperature transport devices and demonstrate nonvolatile switching between distinct quantum phases: paramagnet and ferromagnet.

\section*{Doping-induced Sliding Phase Transition}
We assembled marginally-twisted double bilayer graphene by employing a cut-and-stack method, which yields mixed-domains of rhombohedral and Bernal stacked tetralayer graphene (see Extended Data Fig. 1 for more details). Figure \ref{fig1}b shows a typical real-space image of the domain configuration measured with Kelvin probe force microscopy (KPFM). The surface potential of rhombohedral graphene is $22\ \mathrm{mV}$ smaller than Bernal (Extended Data Fig. 2), namely, the work function of rhombohedral graphene is smaller, consistent with the previous report \cite{atri_spontaneous_2024}. A clear area imbalance is observed between the two domains, with the Bernal domain being larger than the rhombohedral one. This is attributed to the energy difference between the two configurations \cite{li_global_2020, kerelsky_moireless_2021, halbertal_moire_2021}, where Bernal stacking is more energetically favorable at charge neutrality under zero displacement field.

To study the electronic properties of our tetralayer graphene, we fabricated a dual-gated Hall-bar device that enables independent control of the carrier density $n$ and the displacement field $D$. Figure \ref{fig1}c shows the longitudinal resistance $R_{xx}$ as a function of carrier density at a fixed close-to-zero displacement field. We observe an abrupt jump in $R_{xx}$ when doping the graphene from charge neutrality to $n = -15.5 \times 10^{12}\ \mathrm{cm}^{-2}$ (blue curve). This change persists with increasing carrier density (red curve) until it switches back to its original value at $n = -4.1 \times 10^{12}\ \mathrm{cm}^{-2}$. We attribute this hysteretic behavior to a doping-induced sliding phase transition between the Bernal and rhombohedral structures. 

The hole-doped region consists of spatially mixed rhombohedral and Bernal domains, rather than being purely rhombohedral. This interpretation is supported by the absence of clear quantum oscillations in these regions, likely due to the superposition of distinct quantum oscillations originating from the rhombohedral and Bernal domains (Extended Data Fig. 3).  In contrast, the $n > -4.1 \times 10^{12}\ \mathrm{cm}^{-2}$ region appears predominantly Bernal based on the observed quantum oscillations, while it may contain small patches of rhombohedral domains and domain walls (Extended Data Fig. 3) \cite{shi_tunable_2018}. 

The switching between the mixed rhombohedral–Bernal structure (hereafter called ``RB-mixed") and the Bernal-dominant structure (hereafter called ``B-dominant") is driven by the motion of pre-existing domain walls \cite{li_global_2020}, similar to other slidetronic systems \cite{vizner_stern_interfacial_2021, wang_interfacial_2022, weston_interfacial_2022, ko_operando_2023}. Hence, the presence of a mixed-domain structure created by marginal twisting of double bilayer graphene (Fig. \ref{fig1}b) is critical for enabling the switching. We anticipate that a much larger gate voltage would be required to switch a single-domain device, consistent with the absence of gate-induced sliding phase transitions in pure Bernal or rhombohedral tetralayer graphene \cite{shi_tunable_2018, liu_spontaneous_2024, han_signatures_2025}. This also explains why hysteresis appears only in part of the device (see Extended Data Fig. 4), namely, the switching behavior depends on the initial domain configuration and the domain wall pinning strength.

The doping-driven switching can be understood from the difference in the work function between the two structures. As the work function of the rhombohedral structure is smaller, hole doping reduces the Helmholtz free energy difference between the two structures $\Delta F = F_R-F_B$, eventually driving it to change sign. Here, $F_R$ and $F_B$ denote the Helmholtz free energies of the rhombohedral and Bernal structures, respectively. (See Supplementary Information for the detailed discussion) \cite{li_global_2020}. The hole doping enlarges the size of the rhombohedral stacked region, and the switching happens when $\Delta F$ overcomes the certain energy barrier determined by the domain wall pinning strength \cite{vizner_stern_interfacial_2021,ko_operando_2023}. As the switching happens at a constant Free energy difference, it enables us to measure thermodynamic quantities. For example, the magnetic field dependence of the switching boundary allows us to extract the magnetic susceptibility difference between the two structures, which is typically challenging to measure in small-volume 2D devices (Extended Data Fig. 5 and Supplementary Information) \cite{vallejo_bustamante_detection_2021, zhou_imaging_2023}.

The switching is highly robust and reproducible. Figure \ref{fig1}d shows the repeated switching behavior when the sample is subject to a short pulse of gate voltages. This result indicates that the sliding phase transition between Bernal and rhombohedral graphene could be useful as a nonvolatile memory device \cite{yasuda_ultrafast_2024, bian_developing_2024}.

\section*{Phase Diagram}

\begin{figure*}[h]
\centering
\includegraphics[width=1\textwidth]{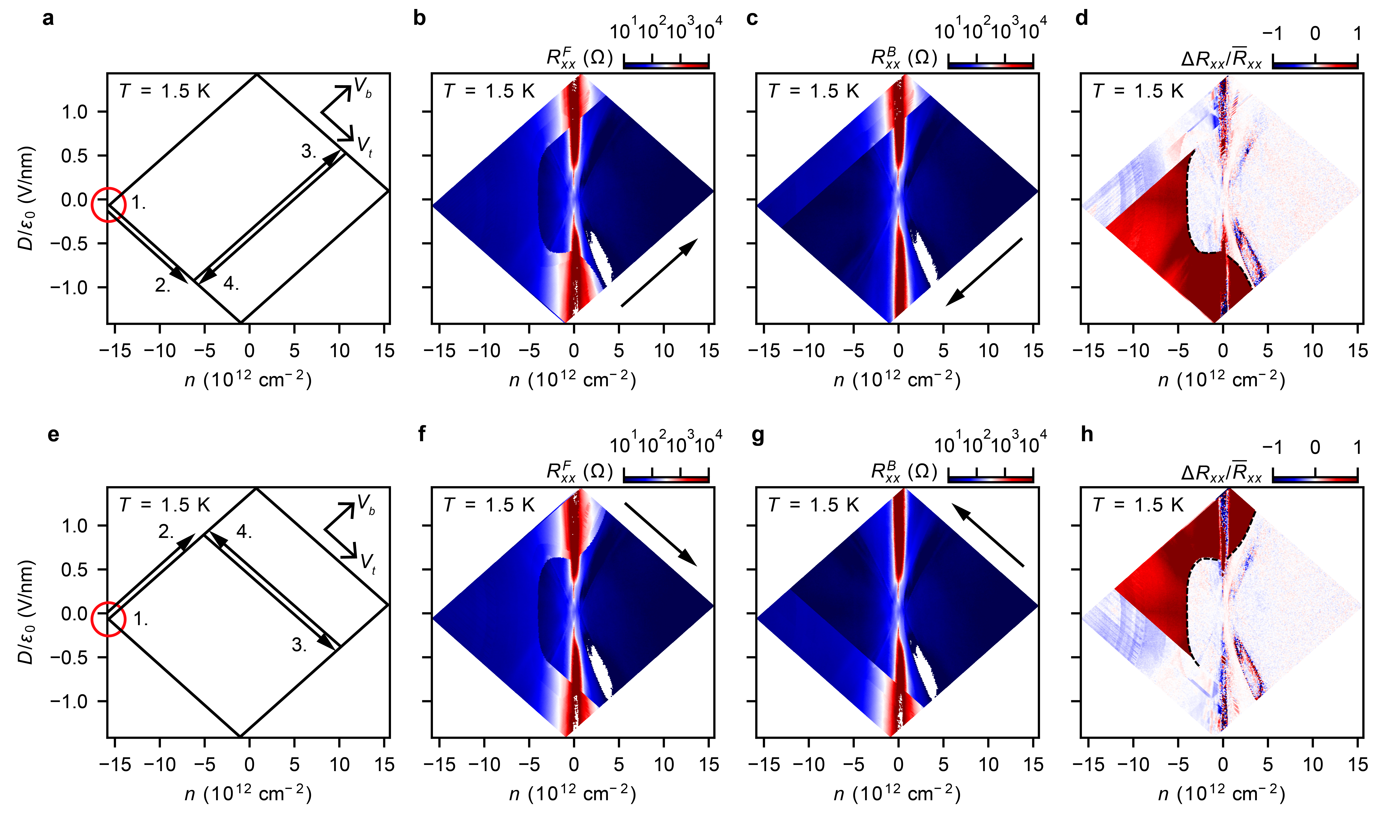}
\caption{\textbf{Determination of switching boundary.} \textbf{a,} Schematic of the measurement procedure for determining the switching boundary between the RB-mixed and B-dominant states. We repeated Steps 1-4 at different values of $V_t$ to obtain \textbf{b} and \textbf{c}. \textbf{b,} $R_{xx}$ measured by sweeping $V_b$ in the increasing direction, $R_{xx}^F$. \textbf{c,} $R_{xx}$ measured by sweeping $V_b$ in the decreasing direction, $R_{xx}^B$. \textbf{d,} The difference, $\Delta R_{xx} = (R_{xx}^F - R_{xx}^B)$, over the average resistance, $\bar{R}_{xx} = (R_{xx}^F + R_{xx}^B) / 2$. The black dashed line corresponds to the switching boundary from RB-mixed to B-dominant state. \textbf{e-h,} The same as \textbf{a-d} with flipped $V_b$ and $V_t$.}
\label{fig2}
\end{figure*}

Having demonstrated carrier-density-induced switching, we next examine the phase diagram as a function of both $n$ and $D$. We follow the measurement procedure outlined in Fig. \ref{fig2}a. First, we dope the sample to the maximum hole density to realize the RB-mixed state (Step 1). Next, we increase the top-gate voltage $V_t$ to a set value while keeping the bottom-gate voltage $V_b$ at its minimum, which preserves the RB-mixed state (Step 2). We then sweep $V_b$ from its minimum to maximum value while measuring the longitudinal resistance $R_{xx}$ (Step 3), and subsequently sweep $V_b$ back from maximum to minimum (Step 4). The structural transitions from RB-mixed to B-dominant happen at certain critical $V_b$ in Step 3. Once the structure becomes B-dominant, it remains B-dominant in Step 4. We repeat Steps 1-4 for different values of $V_t$. Note that the graphene is reset to the RB-mixed state each time it is doped to the maximum hole density in Step 1.

The sets of measurement in Step 3 (Step 4) provide Fig. \ref{fig2}b (Fig. \ref{fig2}c). By subtracting these two scans, we can identify the phase boundary for the RB-mixed-to-B-dominant transition (Fig. \ref{fig2}d) as shown in the black dashed line. We note that $\Delta R_{xx}$ is almost zero around the top left region of the scan (large negative $V_t$ region). This is because the device remains in the RB-mixed state even with the maximum $V_b$ in Step 3. To draw the complete phase boundary, we performed the identical measurement by swapping $V_t$ and $V_b$ in Figs. \ref{fig2}e-h.

\begin{figure*}[h]
\centering
\includegraphics[width=1\textwidth]{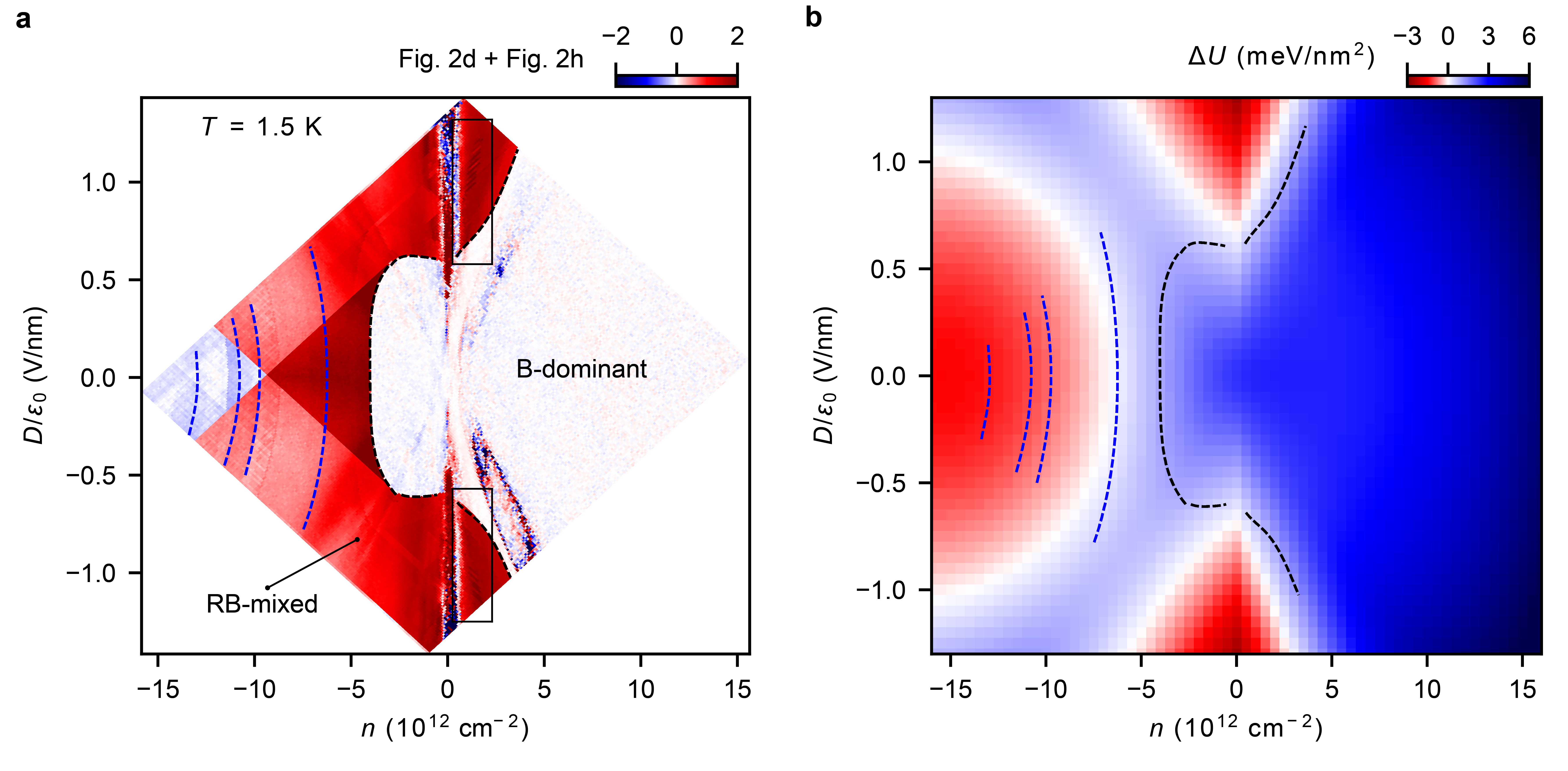}
\caption{\textbf{Phase Diagram. }\textbf{a,} The sum of Fig. \ref{fig2}d and \ref{fig2}h. The left side of the black dashed boundary indicates the area of phase space accessible to the RB-mixed state. The blue dashed lines indicate other equipotential contours that are extracted from another contact configuration (see Extended Data Fig. 7). The black boxes indicate the areas measured in Fig. \ref{fig4}. \textbf{b,} The calculated energy difference, $\Delta U$, between the two structures. Overlaid are the contour lines extracted from the experimental data. }
\label{fig3}
\end{figure*}

Figure \ref{fig3}a combines the two plots from Figs. \ref{fig2}d and \ref{fig2}h to delineate the complete phase diagram. The region to the left of the black dashed boundary—except for the $n < -15.5 \times 10^{12}\ \mathrm{cm}^{-2}$ region—can realize both the RB-mixed and B-dominant states as bistable states, depending on the history of the gate voltage sweep. 

$R_{xx}$ measured using the other side of the Hall bar shows similar resistance jumps (Extended Data Fig. 7), but the magnitude and sign differ from those in Fig. \ref{fig3}a. This contact configuration dependence likely arises from the multi-domain nature of the device, which is a consequence of the marginal twist angle between the two bilayer graphene sheets \cite{yasuda_stacking-engineered_2021,ko_operando_2023}. This structure creates a patchwork of high- and low-resistance regions, leading to a complex current path and contact configuration dependence of the resistance.

We find multiple jumps in the resistance, both in Fig. \ref{fig3}a and Extended Data Fig. 7, in addition to the major jump at the black dashed boundary. This is attributed to the individual motion of the domain walls, which occurs when the energy difference exceeds the energy barrier set by the pinning strength of the domain wall \cite{yasuda_stacking-engineered_2021, ko_operando_2023}. Therefore, each of these jumps corresponds to a line of constant Helmholtz Free energy difference, $\Delta F = const$. Based on this understanding, we can draw constant $\Delta F$ contours on the phase diagram (blue dashed lines). We note that the phase boundaries are mirror symmetric with respect to the line of $D = 0$, which is consistent with the inversion symmetry of both rhombohedral and Bernal structures.

Remarkably, these contour lines agree well with the theoretical calculation in Fig. \ref{fig3}b, which plots the energy difference between the two structures. In addition to the hole-doped region, the large displacement close to the charge neutral regions favors the rhombohedral structure. This originates from the larger gap opening in the rhombohedral graphene as compared to Bernal graphene, which results in a lowering of the total energy of the occupied electrons (see Supplementary Information for the details of calculation and discussion).

\section*{Nonvolatile Control of Magnetism}

\begin{figure*}[h]
\centering
\includegraphics[width=1\textwidth]{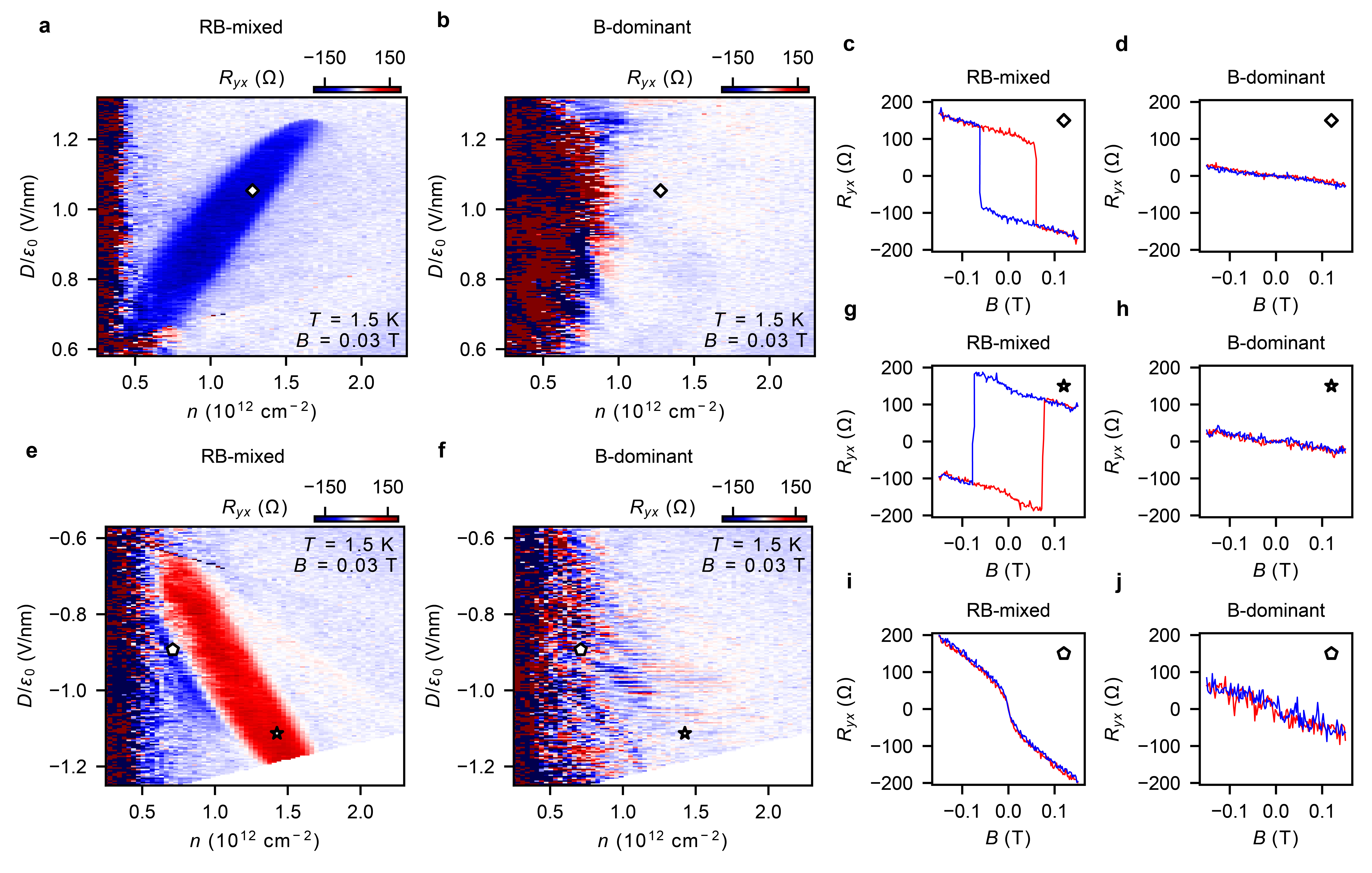}
\caption{\textbf{Nonvolatile control of the anomalous Hall effect. }RB-mixed measurements were taken by first going to the largest hole density corner and then going along the edge of the gate boundary, similar to the measurement in Fig. \ref{fig2}, for each line. B-dominant measurements were taken by first going to $V_t = V_b = 0$ V, before scanning for each line. \textbf{a,}\textbf{b,} $R_{yx}$ measured at $B = 0.03$ T in the positive $D$ and small positive $n$ region (black box in Fig. \ref{fig3}a), in the RB-mixed (\textbf{a}) and B-dominant state (\textbf{b}), respectively. Large negative $R_{yx}$ appears exclusively in the RB-mixed state, indicating an anomalous Hall effect (AHE). \textbf{c,}\textbf{d,} $R_{yx}$ measured at $T=$ 1.5 K as a function of $B$ in both the forward (red) and backward (blue) directions for the RB-mixed (\textbf{c}) and B-dominant state (\textbf{d}), respectively. A nonzero $R_{yx}$ at $B=0$ T and the hysteresis appears only in the RB-mixed state. Note that both of these data are taken at exactly the same gate voltages as highlighted at the diamond position in \textbf{a} and \textbf{b}. \textbf{e,}\textbf{f,} The same as \textbf{a} and \textbf{b} measured in the negative $D$ region. A large positive AHE (and a negative AHE in the smaller $n$ region) appears only in the RB-mixed state. \textbf{g,}\textbf{h,} $R_{yx}$ measured at $T=$ 1.5 K at the star position in the RB-mixed (\textbf{g}) and B-dominant state (\textbf{h}), respectively. \textbf{i,}\textbf{j,} $R_{yx}$ measured at $T=$ 1.5 K at the pentagon position in the RB-mixed (\textbf{i}) and B-dominant state (\textbf{j}), respectively. Nonlinearity in $R_{yx}$ indicates a weak, but finite AHE. }
\label{fig4}
\end{figure*}

Having established the gate-induced sliding phase transition, we next examine its effect on quantum phases in our tetralayer graphene device. We measure the Hall resistance ($R_{yx}$) under a small magnetic field ($B = 0.03$ T) at large positive $D$ and small $n$, corresponding to the region highlighted by the black box in Fig. \ref{fig3}a, for both the RB-mixed state (Fig. \ref{fig4}a) and the B-dominant state (Fig. \ref{fig4}b). In the B-dominant state, we observe a small ordinary Hall effect (OHE) arising from electron doping. In contrast, in the RB-mixed state, we observe a large Hall resistance of $|R_{yx}| > 100 \ \Omega$. Sweeping the magnetic field $B$ in both forward and backward directions reveals a clear magnetic hysteresis (Fig. \ref{fig4}c), namely the anomalous Hall effect (AHE). The temperature-dependent measurement shows the Curie temperature of around 8 K (Extended Data Fig. 8). No such magnetic hysteresis is observed in the B-dominant state, where only OHE is present (Fig. \ref{fig4}d).

Importantly, Figs. \ref{fig4}c and \ref{fig4}d were recorded at exactly the same gate voltage; switching between the two states is achieved solely by the gate-voltage sweep history. This history-dependent bistability demonstrates a nonvolatile transition between ferromagnetic and paramagnetic phases, which is repeatable over many cycles (Extended Data Fig. 9). To the best of our knowledge, this is the first realization of a magnet-to-non-magnet nonvolatile phase transition controllable entirely by electrostatic means \cite{lu_electric-field_2017}.

We perform the same set of measurements in the large negative $D$ and small $n$ region. Similar to the positive $D$ case, we observe a magnet-to-non-magnet phase transition induced by the gate-driven sliding transition (Figs. \ref{fig4}e–h). However, we observe a key distinction: the sign of AHE is reversed compared to the positive $D$ case. This sign reversal of AHE between positive and negative $D$ is surprising, given that both rhombohedral and Bernal graphene are inversion symmetric. In an inversion-symmetric system, the sign of the AHE should remain the same upon reversing $D$. Therefore, the observed AHE sign reversal at the same $n$ but opposite $D$ indicates that the AHE originates from a state that breaks the inversion symmetry.

One possibility is the alignment of hBN, or asymmetries in the device geometry, such as differences in hBN thickness or graphite gate thickness. However, this scenario is unlikely due to the absence of any signatures of moiré superlattice in $R_{xx}$, and the observed mirror symmetry with respect to $D = 0$ in both $R_{xx}$ and the phase boundary (Fig. \ref{fig2}). Furthremore, this sign reversal of AHE is reproducible in another device, namely, this behavior is not specific to the details of the device (Extended Data Figs. 10 and 11).

Therefore, we hypothesize that the sign reversal of the AHE originates from the domain boundary between rhombohedral and Bernal tetralayer graphene. This boundary naturally breaks inversion symmetry, thereby fulfilling the symmetry requirement for the sign reversal of AHE. In addition, comparing with a previous report  \cite{han_signatures_2025}, we find that the region where we observe AHE corresponds to the spin- and valley-polarized quarter-metal phase of pure rhombohedral tetralayer graphene, but with major distinctions where the magnitude of $R_{yx}$ in our device is significantly larger than that reported in pure rhombohedral tetralayer graphene, and a sign reversal is observed as a function of $D$. This observation suggests that the AHE we observe may originate from a proximity effect from the spin- and valley-polarized quarter-metal in the rhombohedral region to either the domain wall or the adjacent Bernal region. Testing this hypothesis will require further experiments, particularly simultaneous real-space imaging of the domain configuration combined with transport measurements \cite{li_global_2020}.

In a small carrier density region of Fig. \ref{fig4}e (around the regions marked by the pentagon), we also observe a negative-sign AHE. This is verified in the magnetic field sweep, where Fig. \ref{fig4}i shows nonlinear behavior in Hall resistance, while being linear in the Bernal structure (Fig. \ref{fig4}j). The AHE origin of nonlinearity is further confirmed through temperature-dependent measurement (Extended Data Fig. 8). Based on the comparisons with a previous report \cite{han_signatures_2025}, we interpret this negative AHE as originating from the quarter-metal phase of the rhombohedral tetralayer graphene region, rather than from the domain boundaries. Although this negative AHE is also likely to exist on the positive $D$ side, the identical sign of the AHE makes it challenging to disentangle the contributions from rhombohedral graphene and domain boundaries. 

In summary, we believe that the observed AHE consists of two components: (i) negative $D$-even AHE in the low–carrier-density region arising from rhombohedral graphene, and (ii) $D$-odd AHE originating from the domain boundary. The understanding of the origin of the latter contribution requires more detailed theoretical and experimental investigations.

\section*{Discussion}

Our research establishes the gate-induced sliding transition as an effective route for controlling quantum phases of matter in 2D systems in a nonvolatile manner. Unlike conventional approaches to structural phase transitions, which often require heating or chemical reactions, this method will allow rapid, energy-efficient, and highly repeatable switching, as demonstrated in other slidetronic systems \cite{yasuda_ultrafast_2024, bian_developing_2024}. The demonstrated reversible on–off control of magnetism through gate-induced sliding phase transition provides a foundation for novel magnetic memory devices and reconfigurable magnetic circuits. We anticipate that our method for introducing a first-order quantum phase transition is applicable to other exotic phases, including chiral superconductivity \cite{han_signatures_2025} and fractional quantum anomalous Hall effect \cite{lu_fractional_2024, xie_tunable_2025}, as well as other 2D materials having metastable stacking orders \cite{li_pressure-controlled_2019,song_switching_2019, klein_enhancement_2019}. Simultaneously, our discovery establishes domain boundaries between the distinct stacking orders of 2D materials as unique and pristine platforms to explore emergent physics \cite{catalan_domain_2012, ju_topological_2015,  chaudhary_superconductivity_2024}. 

\section*{Methods}\label{sec11}

\subsection*{Device Fabrication}

The heterostructure was fabricated using a dry transfer method. First, bilayer graphene, graphite, and hBN flakes were exfoliated onto SiO$_2$/Si substrates, where clean flakes were identified with an optical microscope. A bilayer graphene is cut into two pieces with an anodic oxidation technique  using an atomic force microscope (AFM) (Bruker, Dimension Icon) \cite{masubuchi_fabrication_2009}. A stamp made from polydimethylsiloxane (PDMS, GelPack) covered with poly(bisphenol a carbonate) (PC) was then used to pick up the flakes sequentially with a motorized transfer station (HQ graphene) in a glovebox (MBraun). We first prepared a bottom gate consisting of hBN and graphite, placed it on a substrate, and cleaned it by the AFM in the contact mode. We then picked up graphite and hBN sequentially, followed by a pickup of two pieces of bilayer graphene with nearly-180° or 0.03° rotation, for device A and B, respectively. For device A, one of the layers underwent an additional global shift and relaxed to rhombohedral and Bernal mixed domains (see Extended Data Fig. 1)  \cite{halbertal_unconventional_2022}. This top part of the heterostructure is released to the bottom gate. After the full heterostructure assembly, it was etched into a Hall bar shape using standard electron beam lithography techniques and reactive ion etching (RIE). 1D contacts are made to the device by etching, followed by evaporating Cr/Pd/Au (7/18/40 nm).

\subsection*{Kelvin Probe Force Microscopy Measurement}

KPFM images were acquired using a single-pass frequency modulation technique (FM-KPFM) on an AFM (Bruker, Dimension Icon) with an applied tip bias. Extended Data Fig. 2a was taken while the heterostructure was on the stamp, after two bilayer graphene flakes were picked up. Fig. \ref{fig1}b (identical to Extended Data Fig. 2b) was taken on a heterostructure consisting of twisted double-bilayer graphene/hBN/graphite released on a SiO$_2$/Si substrate. Both twisted double-bilayer graphene were constructed with a rotation angle of nearly 180°, similar to device A. The asymmetry in domain sizes indicates that the sample has relaxed into a mix of rhombohedral and Bernal domains.

\subsection*{Transport Measurement}

All transport measurements were taken in a variable-temperature insert in a $^4$He Cryostat (Oxford Instruments, TeslatronPT). All measurements were taken at 1.5 K, unless otherwise noted. All transport data except Extended Data Figs. 10 and 11 are taken from device A. $R_{xx}$ and $R_{yx}$ were measured with standard 4-probe measurements of the Hall bar using lock-in amplifiers (SRS, SR860) at 17.777 Hz. For device A, we applied a current source of 0.5 nA for the AHE measurements, 1 nA for the measurement in Extended Data Fig. 3c, and 10 nA for all the other measurements. For device B, we used 1 nA for all measurements. We used a current preamp (Basel Precision Instruments, SP983c-IF) and voltage preamps (Basel Precision Instruments, SP1004) to reduce noise. The top and bottom gate voltages $V_t$ and $V_b$ are supplied by sourcemeters (Keithly, 2450). Carrier density and displacement field are defined as $n = (C_t V_t + C_b V_b)/e - n_0$ and $D = (-C_t V_t + C_b V_b)/2 - D_0$. $C_t$ and $C_b$ are the top-gate and bottom-gate capacitance per area, and $n_0$ and $D_0$ are the offset carrier density and displacement field at $V_t = V_b = 0$ V. These values are calibrated from the Landau levels in the B-dominant state. We anti-symmetrized $R_{yx}$ as a function of $B$ to remove any leakover of $R_{xx}$ due to contact misalignment. 

\subsection*{Calculation}

In the Supplementary Information, we include the details of the calculation of the energy difference $\Delta U$ between Bernal-stacked tetralayer graphene and rhombohedral-stacked tetralayer graphene. In short, by integrating over the tight-binding band structure, we can determine the energy difference between the two states at any given $n$ and $D$. Besides tight-binding parameters, this leaves only two free parameters, the work function difference $\Delta \Phi$ and the stacking energy difference $\Delta U (0,0)$ at $n = D = 0$. We extracted $\Delta \Phi$ from our KPFM (Extended Data Fig. 2). The determination of $\Delta U (0,0)$ is explained in the Supplementary Information. 

\bmhead{Acknowledgements}

We acknowledge the helpful discussions with Yawei Zhang and James C. M. Hwang, and experimental support by Austin Wu and Junhao Lin. 
This study was supported by a startup fund at Cornell University. 
This work was performed in part at the Cornell NanoScale Facility, a member of the National Nanotechnology Coordinated Infrastructure (NNCI), which is supported by the National Science Foundation (Grant NNCI-2025233). 
This work made use of the Cornell Center for Materials Research shared instrumentation facility. 
T.D. acknowledges support from a startup fund at Stanford University.
K.W. and T.T. acknowledge support from the JSPS KAKENHI (Grant Numbers 21H05233 and 23H02052) , the CREST (JPMJCR24A5), JST and World Premier International Research Center Initiative (WPI), MEXT, Japan.

\bmhead{Contributions}
K.Y. conceived and supervised the project. D.B., Y.A., P.G. and A.G. fabricated and imaged the device, and performed transport measurement with the help of L.M.. T.T. and T.D. performed the theoretical calculations. K.W. and T.T. grew hBN crystals. D.B. and K.Y. wrote the manuscript from contributions from all authors.

\bmhead{Competing interests}
The authors declare no competing interests.

\bibliography{references,references_Tixuan,references_Daniel}

\end{document}